\documentclass[aps,preprint]{revtex4}%
\usepackage{amsfonts}
\usepackage{amsmath}
\usepackage{amssymb}
\usepackage{graphicx}%
\begin{document}
\title{Atomic oxygen adsorption and incipient oxidation of the Pb(111) surface: A
density-functional theory study }
\author{Bo Sun}
\affiliation{Institute of Applied Physics and Computational Mathematics, P.O. Box 8009,
Beijing 100088, People's Republic of China}
\author{Ping Zhang}
\thanks{Corresponding author. Electronic address: zhang\_ping@iapcm.ac.cn}
\affiliation{Institute of Applied Physics and Computational
Mathematics, P.O. Box 8009, Beijing 100088, People's Republic of
China}
\author{Zhigang Wang}
\affiliation{Institute of Applied Physics and Computational
Mathematics, P.O. Box 8009, Beijing 100088, People's Republic of
China}
\author{Suqing Duan}
\affiliation{Institute of Applied Physics and Computational
Mathematics, P.O. Box 8009, Beijing 100088, People's Republic of
China}
\author{Xian-Geng Zhao}
\affiliation{Institute of Applied Physics and Computational
Mathematics, P.O. Box 8009, Beijing 100088, People's Republic of
China}
\author{Xuchun Ma and Qi-Kun Xue}
\affiliation{Institute of Physics, Chinese Academy of Sciences, Beijing 100080, People's
Republic of China}

\begin{abstract}
We study the atomic oxygen adsorption on Pb(111) surface by using
density-functional theory within the generalized gradient approximation and a
supercell approach. The atomic and energetic properties of purely on-surface
and subsurface oxygen structures at the Pb(111) surface are systematically
investigated for a wide range of coverages and adsorption sites. The fcc and
tetra-II sites (see the text for definition) are found to be energetically
preferred for the on-surface and subsurface adsorption, respectively, in the
whole range of coverage considered. The on-surface and subsurface oxygen
binding energies monotonically increase with the coverage, and the latter is
always higher than the former, thus indicating the tendency to the formation
of oxygen islands (clusters) and the higher stability of subsurface
adsorption. The on-surface and subsurface diffusion-path energetics of atomic
oxygen, and the activation barriers for the O penetration from the on-surface
to the subsurface sites are presented at low and high coverages. In
particular, it is shown that the penetration barrier from the on-surface hcp
to the subsurface tetra-I site is as small as 65 meV at low coverage ($\Theta
$=0.25). The other properties of the O/Pb(111) system, including the charge
distribution, the lattice relaxation, the work function, and the electronic
density of states, are also studied and discussed in detail, which
consistently show the gradually stabilizing ionic O-Pb bond with increase of
the oxygen coverage.

\end{abstract}
\maketitle

\section{INTRODUCTION}

Lead (Pb) is one kind of heavy metal element with widespread availability.
Single crystal Pb is extremely resistant to oxygen \cite{Pb}. Appreciable
oxidation takes place only at relatively high temperature ($\geqslant$370K) or
at considerable oxygen coverage at room temperature. However, this
inoxidability of lead can be overcome by a so-called two-step treatment
\cite{Jiang1}: (i) low temperature (100K) adsorption, followed by (ii)
annealing at elevated temperatures to above 220K. The method itself proves to
be a very effective way for low-temperature surface oxidation. Using this
two-step approach, recently, the oxygen adsorption on ultra-thin Pb(111) films
was experimentally investigated by using the scanning tunneling microscopy
(STM) and scanning tunneling spectroscopy (STS) measurements \cite{PNAS}.
Remarkably, it was found that the surface oxidation displays a prominent
quantum oscillating effect by varying the thickness of Pb(111) film.
Furthermore, it was found that the O adsorbates can form some magic clusters
with regular size and shape \cite{Jiang2}. These phenomena of the oxygen
adsorption and surface oxidation at Pb(111) remain yet to be exploited and
understood, which is a main driving force for our present first-principles
study. Contrary to the extensive first-principles studies of the surface
oxidation on transition metals (TMs) such as Cu
\cite{Cu(100),Cu(100)-1,Cu(100)-2,Cu(111)-1,Cu(111)-2,Cu(111)-3,Cu(111)-4,Cu(111)-5,Cu(111)-6}%
, \ Ag \cite{Ag(001),Ag(001)-1,Ag(110),Ag(110)-1,Ag(111)-00,Ag(111)-0,
LiWX-2002, LiWX-2003, Ag(111)-1,Ag(111)-2,Ag(111)-3,Ag(111)-4}, Rh
\cite{Rh(111)-1,Rh(111)-2,Rh(111)-3,Rh(111)-4,Rh(111)-5}, Pd
\cite{Intro-14,Pd(111)-1,Pd(111)-2,Pd(111)-3,Pd(111)-4}, Pt
\cite{Pt(110),Pt(111)-0,Pt(111)-1,Pt(111)-2,Pt(111)-3}, and Ru
\cite{Intro-11,Ru-1}, and on the simple $sp$ metals such as Al
\cite{Intro-5,Intro-6,Intro-7,Intro-8,Al(111)-1,Al(111)-2,Al(111)-3} and Mg
\cite{Intro-9,Intro-12,Mg(0001)}, \textit{ab initio} studies of oxygen
chemisorption on Pb(111) surface, in particular, O subsurface species on
Pb(111) surface, are still lacking.

In this paper, we have carried out the first-principles calculations of the
on-surface and the subsurface O adsorption at Pb(111) in a wide range of
coverages. Results for the determination of stable adsorption sites,
work-function and atomic-relaxation changes, charge densities, electronic
structures, and energy barriers for O diffusion and penetration, are
systematically presented. Like the other metals, the surface oxidation of Pb
is expected to involve three main events: (i) the initial dissociation of
O$_{2}$ molecules and the oxygen chemisorption on Pb(111) surface, followed by
(ii) lattice penetration of atomic oxygen and oxide nucleation, and finally
(iii) the crystallization and growth of the stoichiometric oxide phases. The
sequence of these events may be complex, and rather than successively, they
may occur simultaneously, depending on the temperature and pressure. The main
purpose of this paper is to give a partial but detailed understanding of
events (i) and (ii) by studying the energetics and structures of atomic O
adsorbates with purely on-surface and subsurface adsorption sites and with
different coverages. A full study involving the initial dissociation of
O$_{2}$ molecules when approaching to or accumulating on the Pb(111) surface,
and the penetration process which involves simultaneous on-surface and
sub-surface adsorption, will be given elsewhere. In particular, since the
atomic oxygen adsorption and diffusion on metal surface are elementary
processes towards the whole surface oxidation, and the atomic configuration
after the oxygen chemisorption may give a reasonable foresee for the tendency
of bulk oxidation, it is expected that the present systematic first-principles
calculations of the oxygen adsorption on Pb(111) is of highly interest in
relation to understanding the nature of the O-Pb bond in general and of great necessity.

This paper is organized as follows. In Sec. II\ we give details of the
first-principles total energy calculations, which is followed in Sec. III by
our results for bulk Pb, the clean Pb(111) surface, and the O$_{2}$ molecule.
The results for purely on-surface adsorption as a function of the oxygen
coverage are presented in Sec. IV, where the surface and adsorption
energetics, the atomic geometry, and the electronic structures are presented
and analyzed. In Sec. V, we discuss the purely subsurface adsorption of oxygen
atoms as we did in Sec. IV. The energy barriers for atomic oxygen diffusion
and penetration are presented in Sec. VI, and the conclusion is given in Sec. VII.

\section{THE CALCULATION METHOD}

The density-functional theory (DFT) total energy calculations were carried out
using the Vienna \textit{ab initio} simulation package \cite{Vasp} with the
projector-augmented-wave (PAW) pseudopotentials \cite{Paw} and plane waves
\cite{Plane Wave}. In the present film calculations, the so-called
\textit{repeated slab} geometries were employed \cite{Slab}. This scheme
consists in the construction of a unit cell of an arbitrarily fixed number of
atomic layers identical to that of the bulk in the plane of the surface
(defining the bi-dimensional cell), but symmetrically terminated by an
arbitrarily fixed number of empty layers (the \textquotedblleft\textit{vacuum}%
\textquotedblright) along the direction perpendicular to the surface. In the
present study, the clean Pb(111) surface is modeled by periodic slabs
consisting of ten lead layers separated by a vacuum of 20 \r{A}, which is
found to be sufficiently convergent. The oxygen atoms are adsorbed on both
sides of the slab in a symmetric way. During our calculations, the positions
of the outmost three lead layers as well as the O atoms are allowed to relax
while the central four layers of the slab are fixed in their calculated bulk
positions. The plane-wave energy cutoff was set 400 eV. If not mentioned
differently we have used a $(12\times12\times1)$ $k$-point grid for the
p$(1\times1)$ surface cell, $(6\times6\times1)$ $k$-point grid for the
p$(\sqrt{3}\times\sqrt{3})$ and p$(2\times2)$ cells, and $(4\times4\times1)$
$k$-point grid for the p$(3\times3)$ cell, with Monkhorse-Pack scheme
\cite{Pack}. The use of larger $k$-point meshes did not alter these values
significantly. Furthermore, the generalized gradient approximation (GGA) of
Perdew \textit{et al}. \cite{GGA-2} for the exchange-correlation potential was
employed. A Fermi broadening \cite{Fermi Broaden} of 0.05 eV was chosen to
smear the occupation of the bands around $E_{F}$ by a finite-$T$ Fermi
function and extrapolating to $T=0$ K.%

\begin{figure}[tbp]
\begin{center}
\includegraphics[width=0.6\linewidth]{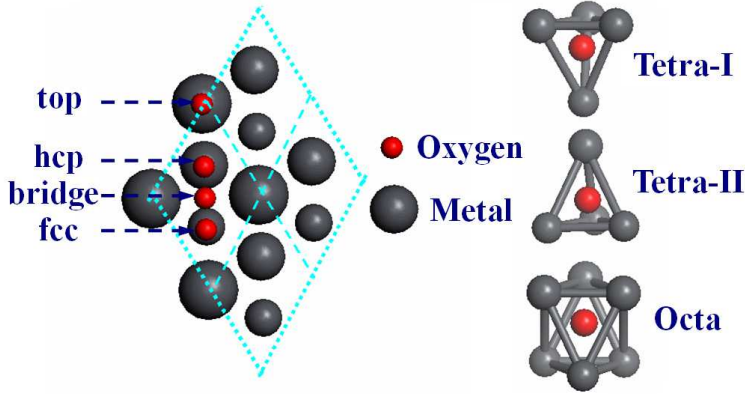}
\end{center}
\caption
{(Color online) (Left panel) four on-surface adsorption sites including fcc, hcp, bridge (B2), and on-top (T1) sites. Note that Pb atoms of outmost three layers are shown by scaled grey balls
(the largers are outer Pb atoms). (Right panel) three subsurface adsorption sites including tetra-I, tetra-II, and octa sites.}
\label{fig1}
\end{figure}%
In the present paper, the calculations for oxygen atoms in the five adsorption
sites, including on-surface (hcp and fcc) and subsurface (tetra-I, tetra-II,
and octa) sites depicted in Fig. 1, have been performed for coverage ranging
from 0.11 ML to a full monolayer. Specially, the oxygen coverages of 0.11 ML,
0.33 ML, and 0.67 ML were calculated using p$(3\times3)$\ surface unit cell,
while the coverages of 0.25 ML, 0.50 ML, 0.75 ML, and 1.0 ML\ were calculated
in the p$(2\times2)$ surface cell containing one, two, three, and four oxygen
atoms, respectively. The on-surface top (T1) and the bridge (B2) adsorption
sites were also considered. The T1 site was found to be notably less favorable
than the fcc and hcp sites. When the O atom is placed on the B2 site, it
always moves to the fcc site after relaxation. Actually, figure 15 below will
show that the B2 site is a saddle point in the O diffusion path from hcp to
fcc site. Thus in this paper, most of the on-surface adsorption studies are
focused on the fcc and hcp sites.

One central quantity tailored for the present study is the average binding
energy of the adsorbed oxygen atom defined as
\begin{equation}
E_{b}(\Theta)=-\frac{1}{N_{\text{O}}}[E_{\text{O/Pb(111)}}-E_{\text{Pb(111)}%
}-N_{\text{O}}E_{\text{O}}], \tag{2}%
\end{equation}
where $N_{\text{O}}$ is the total number of O atoms (on-surface and
subsurface) present in the supercell at the considered coverage $\Theta$ (we
define $\Theta$ as the ratio of the number of adsorbed atoms to the number of
atoms in an ideal substrate layer). $E_{\text{O/Pb(111)}}$, $E_{\text{Pb(111)}%
}$, and $E_{\text{O}}$\ are the total energies of the slab containing oxygen,
of the corresponding clean Pb(111) slab, and of a free O atom respectively.
According to this definition, $E_{b}$ is also the adsorption energy
$E_{\text{ad}}$ per O atom, i.e., the energy that a free O atom gains upon its
adsorption. Thus a positive value of $E_{b}$ indicates that the adsorption is
exothermic (stable) with respect to a free O atom and a negative value
indicates endothermic (unstable) reaction. On the other hand, since in most
cases, the oxygen chemisorption process inevitably involves the dissociation
of O$_{2}$ molecules, thus the adsorption energy per oxygen atom can
alternatively be referenced to the energy which the O atom has in the O$_{2}$
molecule by subtracting half the dissociation energy $D$ of the O$_{2}$
molecule,
\begin{equation}
E_{\text{ad(}1/2\text{O}_{2}\text{)}}=E_{b}-D/2. \tag{3}%
\end{equation}
With this choice of adsorption energy, then a positive value indicates that
the dissociative adsorption of O$_{2}$\ is an exothermic process, while a
negative value indicates that it is endothermic and that it is energetically
more favorable for oxygen to be in the gas phase as O$_{2}$. \ \

\section{\bigskip THE BULK Pb, CLEAN Pb(111) SURFACE, AND OXYGEN MOLECULE}

First the total energy of the bulk fcc Pb was calculated to obtain the bulk
lattice constant. The calculated lattice constant is $a$=5.028 \r{A},
comparable well with experimental values of 4.95 \r{A}\cite{Kittel}. The 6$s$
and 6$p$ electrons of the Pb atom were treated as the valence electrons and
the 5$d$ electrons were treated as core electrons. This choice of valence and
core electrons has been previously employed to study the electronic properties
of lead oxide (including PbO \cite{PbO} and PbO$_{2}$ \cite{PbO2}). Note that
we have also studied the effect of Pb 5d electrons on the surface properties
of clean and oxygen-adsorbed Pb(111) surface and found no accountable changes
by treating the Pb 5d electrons as valence electrons \cite{Note1}. The
orbital-resolved electronic density of states (DOS) per atom for the bulk Pb
is shown in Fig. 2(a). The two broad peaks correspond to Pb 6$s$ and 6$p$
states. Compared to the 6$s$ band, the 6$p$ bands are more dispersive due to
the extensive character of the atomic $p$ orbitals. In addition, there is a
small $s$-$p$ hybridization near the Fermi energy.%

\begin{figure}[tbp]
\begin{center}
\includegraphics[width=1.0\linewidth]{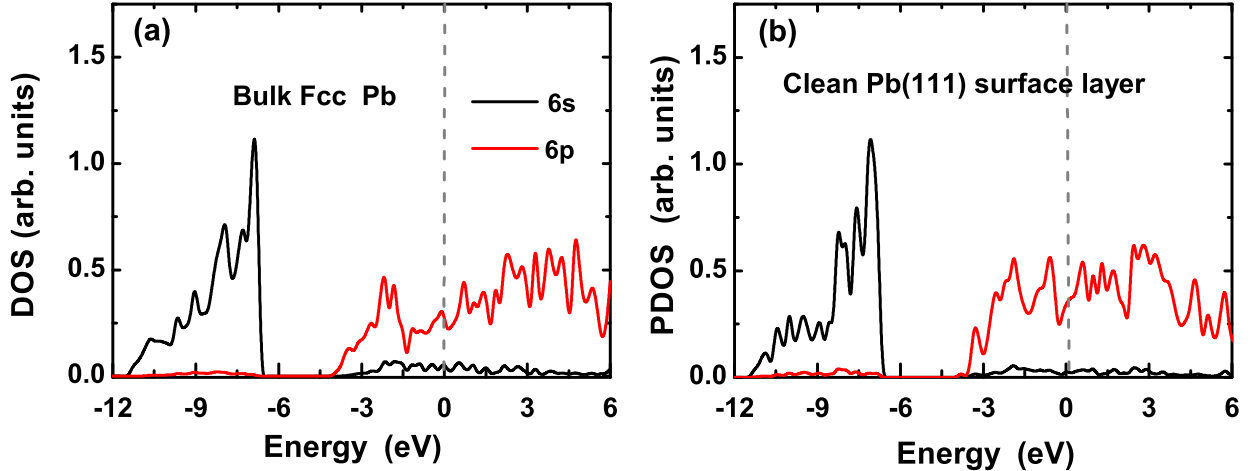}
\end{center}
\caption
{(Color online) (a) Orbital-resolved DOS (per atom) for bulk Pb. (b) Orbital-resolved
partial DOS for the clean $1\times
1$ Pb(111) surface layer. The Fermi energy is indicated by the vertical dashed line at 0 eV.}
\end{figure}%
The calculation for the atomic relaxations of the clean surface p$(1\times1)$,
p$(\sqrt{3}\times\sqrt{3})$, p$(2\times2)$, and p$(3\times3)$\ periodicities
(models), provides not only a test of the clean surface with different cell
sizes, but is also used to evaluate the charge density difference used later
and assess the changes in the work function by the O adsorption. Our
clean-surface calculations show that the two outmost Pb(111) layers relax
significantly from the bulk values, namely, the first-second layer contraction
is nearly 5\% and the second-third layer expansion is nearly 3\%, in agreement
with the recent results from first-principles calculations \cite{Wu Biao} and
the LEED experiment \cite{LEED}. Note that the first interlayer separation on
most metal surfaces is contracted, Pb(111) is also a typical example. The
third-forth layer distance is practically the same as the interlayer distance
in the bulk. The variation of the work function is negligible for different
surface cells [with a value of 3.83 eV, see the inset in Fig. 7(a) below] and
is in excellent agreement with the other theoretical result \cite{Wu Biao}.
The charge density $n(\mathbf{r})$ (not depicted here) shows that like the
other typical surface calculations \cite{Al001}, there is a rapid variation in
$n(\mathbf{r})$ in the surface interstitial region, with $n(\mathbf{r})$
falling off sharply in magnitude towards the vacuum and soon \textquotedblleft
healing\textquotedblright\ the discrete atomic nature of the surface. This
sizable charge redistribution near the surface is associated with the
formation of the (uniform) surface dipole layer, which sensitively determines
the work function.

As one knows, surface calculations are very subtle, requiring enough $k$-point
mesh and efficient energy cutoff, the correct model, and the other details. To
test the convergency of the physical properties of the clean Pb(111) surface,
we have calculated and analyzed the interlayer relaxations $\Delta_{ij}%
$=$(d_{ij}-d_{0})/d_{0}$ with respect to the bulk interlayer distance $d_{0}%
$=2.903 \AA , the surface energy $E_{s}$, and the work function $\Phi$ of the
clean Pb(111) films by using different models with various $k$-points mesh.
The results for different models are listed in Table I, from which it reveals
that while the atomic interlayer relaxations are somewhat sensitive to the
choice of $k$-points mesh and the model, the influences to the surface
energetics from using different models are negligibly small. We suggest to
analyze the surface energetics and electronic structures under the
calculations of the same model, since the required high accuracy is expected
to obtain by using the same surface model to do more reliable calculations and
comparisons. For example, a p$(3\times3)$ surface unit cell is needed if one
wants to analyze the properties of surface chemical activity\ in a coverage
range beginning\ from $\Theta$=0.11. \begin{table}[th]
\caption{The calculated interlayer relaxation $\Delta_{ij}$ (\%), surface
energy $E_{s}$ (in eV), and work function $\Phi$ (in eV) for different clean
Pb(111) surface models with different $k$-point meshes. }
\begin{tabular}
[c]{cccccccc}\hline\hline
Model & $k$-point mesh & irreducible $k$ & $\Delta_{12}$ (\%) & $\Delta_{23}$
(\%) & $\Delta_{34}$ (\%) & $E_{s}$ (eV) & $\Phi$ (eV)\\\hline
$1\times1$ & $12\times12\times1$ & 216 & -4.670 & 2.560 & -0.149 & 0.372 &
3.830\\
\ $1\times1$ & $6\times6\times1$ & 54 & -5.326 & 3.066 & -0.267 & 0.350 &
3.786\\
\ $\sqrt{3}\times\sqrt{3}$ & $6\times6\times1$ & 54 & -5.110 & 2.090 & 0.160 &
0.360 & 3.833\\
\ $2\times2$ & $6\times6\times1$ & 54 & -4.803 & 2.977 & -0.100 & 0.372 &
3.830\\
\ $3\times3$ & $4\times4\times1$ & 24 & -4.882 & 2.890 & 0.369 & 0.373 &
3.832\\
\ $3\times3$ & $3\times3\times1$ & 5 & -5.978 & 2.831 & 1.320 & 0.372 &
3.834\\\hline\hline
\end{tabular}
\end{table}

Figure 2(b) plots the orbital-resolved layer-projected density of states
(PDOS) for the topmost Pb layer of the clean p$(1\times1)$\ Pb(111) surface
cell. Compared to the Fig. 2(a), one can see that the surface electronic
structure is almost the same as that of the bulk, with a little anisotropy in
the $p_{x}/p_{y}$ and $p_{z}$ orbitals for the surface Pb atom. Note that
throughout this paper we did not consider the quantum size effect on the
atomic and electronic structures, since in our various supercell models the
substrate has been fixed with the same thickness.

The total energies of the isolated O atom and the free O$_{2}$ molecule are
calculated in a cubic cell of side length 10 \r{A}with a $(3\times3\times
3)$\ $k$-point mesh for the Brillouin zone sampling. The spin-polarization
corrections to the O atom and the O$_{2}$ molecule are included. The binding
energy of O$_{2}$\ is calculated to be $1/2E_{b}^{\text{O}_{2}}$=3.12\ eV per
atom and the O-O bond length is about 1.235 \r{A}.\ These results are typical
for well-converged DFT-GGA calculations. Compared to the experimental
\cite{Molecule Expt} values of $2.56$\ eV\ and 1.21 \r{A}for O binding energy
and bonding length, the usual DFT-GGA result always introduces an
overestimation, which reflects the theoretical deficiency for describing the
local orbitals of the oxygen. The resultant error in calculating the absolute
value of the binding energy, however, does not matter in this work, since it
is the difference in binding energies of two geometries that determines which
one is more sable (if they contain the same amount of oxygen) or how $E_{b}$
evolves with coverage (if the structures contain an unequal amount of O
atoms). We will consider this overbinding of O$_{2}$\ when drawing any
conclusion that may be affected by its explicit value.

\section{THE PURE ON-SURFACE ADSORPTION}%

\begin{figure}[tbp]
\begin{center}
\includegraphics[width=0.7\linewidth]{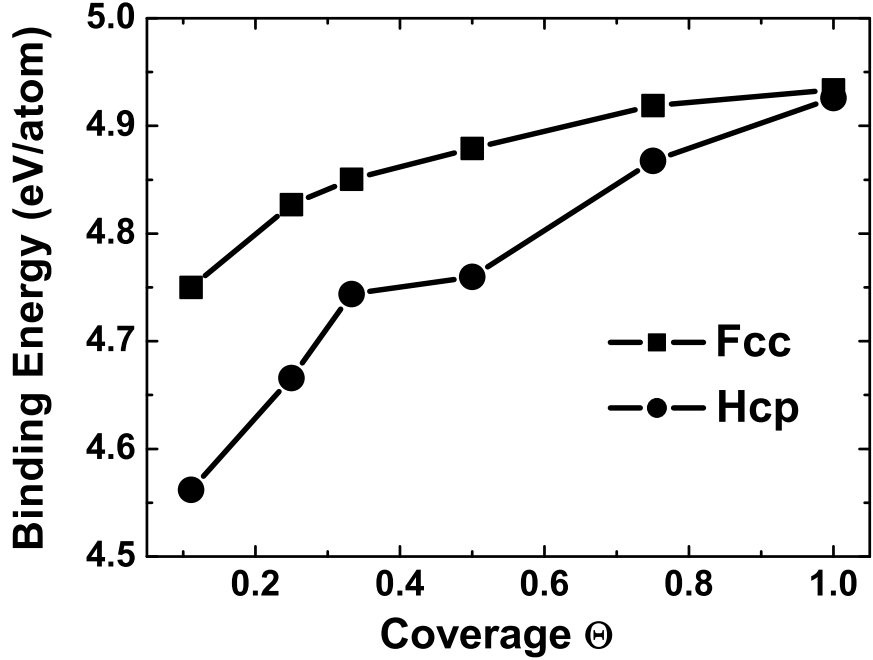}
\end{center}
\caption{Calculated binding energies of O/Pb(111)
systems, with oxygen in the fcc (solid squares) and hcp (solid circles)
sites, as functions of the oxygen coverage. The solid lines connecting
the calculated binding energies are used to guide the eyes.}
\end{figure}
\begin{table}[th]
\caption{The calculated binding energy $E_{b}$ (in eV) and work function
$\Phi$ (in eV) as functions of oxygen\ coverage for five on-surface and
subsurface adsorption sites. }
\begin{tabular}
[c]{cccccccc}\hline\hline
& Site & $\Theta$=$0.11$, \  & 0.25, \  & 0.33, \  & 0.5, \  & 0.75, \  & 1.0
\ \\\hline
& fcc & 4.750 \  & 4.827 \  & 4.850 \  & 4.879 \  & 4.919 \  & 4.934 \ \\
& hcp & 4.562 \  & 4.666 \  & 4.744 \  & 4.760 \  & 4.868 \  & 4.926 \ \\
E$_{b}$ \ \  & tetra-I & 4.771 \  & 4.785 \  & 4.863 \  & 4.887 \  & 5.019
\  & 5.139 \ \\
& tetra-II & 4.842 \  & 4.907 \  & 4.927 \  & 4.957 \  & 5.053 \  & 5.145 \ \\
& octa & 4.628 \  & 4.545 \  & 4.776 \  & 4.876 \  & 4.862 \  & 5.012
\ \\\hline
& fcc & 3.941 \  & 4.222 \  & 4.407 \  & 4.715 \  & 5.269 \  & 5.969 \ \\
& hcp & 3.966 \  & 4.253 \  & 4.485 \  & 4.749 \  & 5.236 \  & 5.994 \ \\
$\Phi$ \ \ \  & tetra-I & 3.812 \  & 3.759 \  & 3.736 \  & 3.499 \  & 3.490
\  & 3.148 \ \\
& tetra-II & 3.809 \  & 3.706 \  & 3.621 \  & 3.543 \  & 3.289 \  & 2.997 \ \\
& octa & 3.807 \  & 3.749 \  & 3.616 \  & 3.567 \  & 3.322 \  & 3.062
\ \\\hline\hline
\end{tabular}
\end{table}For different oxygen coverages $\Theta$, the binding energies
$E_{b}$ for oxygen on the Pb(111) surface in the fcc and hcp sites, with
respect to atomic oxygen, are illustrated in Fig. 3 and summarized in Table
II. One can see from Fig. 3 that the binding energy for the O$_{\text{fcc}}$
or O$_{\text{hcp}}$ atom displays a modestly\ increasing tendency with the
oxygen coverage, while the overall variation in the magnitude of $E_{b}$ is
rather small in the range of coverage we considered. The increasing binding
energy with coverage indicates a prominent attraction between the on-surface
oxygen atoms, implying a tendency to form the oxygen islands or clusters on
the Pb(111) surface. This result is similar to that of the O-adsorbed Al(111)
and Mg(0001) surfaces, compared to which the increasing slope in $E_{b}%
$-$\Theta$ curve in the present O/Pb(111) system is somewhat slower, but is in
contrast to the oxygen/TMs due to the increasing electron occupation of the
antibonding states with the O coverage in those systems. On the other side,
considering the binding energy $E_{b}$ of atomic oxygen on the surface with
respect to half the binding energy of the O$_{2}$ molecule (the calculated
value of 3.12 eV or experimental value of 2.56 eV), we can see that the
calculations predict that in the whole coverage range we considered, the
atomic oxygen on-surface adsorption is stable. Also it reveals in Fig. 3 that
the fcc site is energetically favorable relative to the hcp site, although
their difference in $E_{b}$ decreases with increasing $\Theta$, namely, from
0.19 eV at $\Theta$=0.11 down to (less than) 0.01 eV at $\Theta$=1.0.

\begin{table}[th]
\caption{The calculated atomic interlayer relaxation $\Delta_{ij}$ (\%), O-Pb
bond length $R_{1}$ (in \AA ), and the adsorbate height $h_{\text{O-Pb}}$ (in
\AA ) for different oxygen\ coverages of on-surface adsorption.}
\begin{tabular}
[c]{ccccccccc}\hline\hline
Coverage & \multicolumn{2}{c}{$h_{\text{O-Pb}}$ (\AA ) \ } &
\multicolumn{2}{c}{$R_{1}$ (\AA ) \ } & \multicolumn{2}{c}{$\Delta_{12}$
($\%$) \ } & \multicolumn{2}{c}{$\Delta_{23}$ ($\%$)}\\
& \multicolumn{2}{c}{fcc \ \ hcp \ } & \multicolumn{2}{c}{fcc \ \ hcp \ } &
\multicolumn{2}{c}{fcc \ \ hcp \ } & \multicolumn{2}{c}{fcc \ \ hcp
\ }\\\hline
0.11 & 1.116 & 1.115 \  & 2.282 & 2.289 \  & -3.098 & -3.567 \  & 1.752 &
-0.492 \ \\
0.25 & 0.914 & 0.876 \  & 2.266 & 2.268 \  & -1.641 & -0.413 \  & 1.378 &
-0.601 \ \\
0.33\  & 1.047 & 0.939 \  & 2.257 & 2.274 \  & -1.568 & -0.095 \  & 0.779 &
-0.176 \ \\
0.50 & 0.975 & 1.010 \  & 2.265 & 2.269 \  & 3.405 & 4.562 \  & -0.614 &
-2.498 \ \\
0.75 & 0.980 & 0.944 \  & 2.258 & 2.252 \  & 12.995 & 12.298 \  & -2.445 &
-3.243 \ \\
1.00 & 0.971 & 0.970 \  & 2.271 & 2.272 \  & 20.363 & 21.002 \  & -5.085 &
-4.863 \ \\\hline\hline
\end{tabular}
\end{table}Table III presents the results for the relaxed atomic structures,
including the height $h_{\text{O-Pb}}$ of O above the surface, the O-Pb bond
length $R_{1}$, and the interlayer relaxations $\Delta_{ij}$ for various
coverages with O in the fcc and hcp sites.
\begin{figure}[tbp]
\begin{center}
\includegraphics[width=0.7\linewidth]{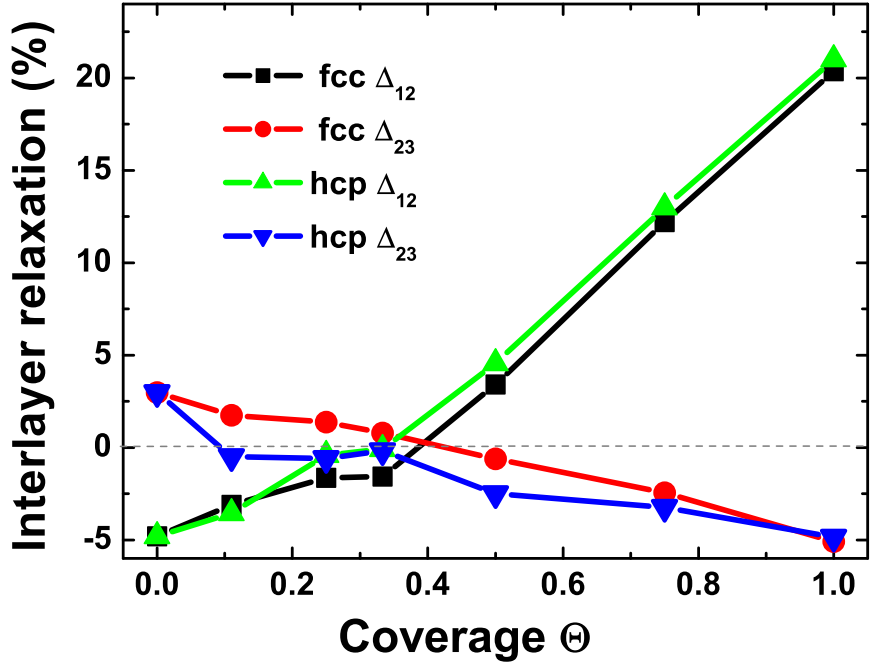}
\end{center}
\caption{(Color online) Atomic interlayer relaxations as
functions of coverage for oxygen in fcc (solid square)
and hcp (solid circle) sites.}
\end{figure}%
For more clear illustration, the first and second interlayer relaxations
($\Delta_{12}$ and $\Delta_{23}$), and the O-Pb bond length are also plotted
in Fig. 4 and Fig. 5, respectively. One can see from Fig. 4 that the
adsorption of oxygen on Pb(111) induces notable changes in the interlayer
distances of the substrate for the whole coverages considered. In particular,
the topmost interlayer relaxation ($\Delta_{12}$) changes from contraction
(about $-$5\%)\ to expansion (about 20\%), and on the contrary, the second
interlayer relaxation ($\Delta_{23}$) changes from expansion (about 3\%)\ to
contraction (about $-$5\%), for O in both fcc and hcp sites. This large and
sign-inverted change of atomic interlayer relaxation by oxygen adsorption is
unique for O/Pb(111) when comparing with the other oxygen/metal systems. For
instance, at Al(111) surface, the topmost interlayer relaxation on oxygen
adsorption is at most $\Delta_{12}$=$4\%$ \cite{Al(111)-2}, while at Ag(111)
surface, there is even no change of interlayer relaxation before and after
oxygen adsorption. The large expansion in Pb(111) interlayer relaxations
reflects the strong influence of the O adsorbate on the neighboring Pb atoms,
and will result in important redistribution of the electronic structure.%

\begin{figure}[tbp]
\begin{center}
\includegraphics[width=0.7\linewidth]{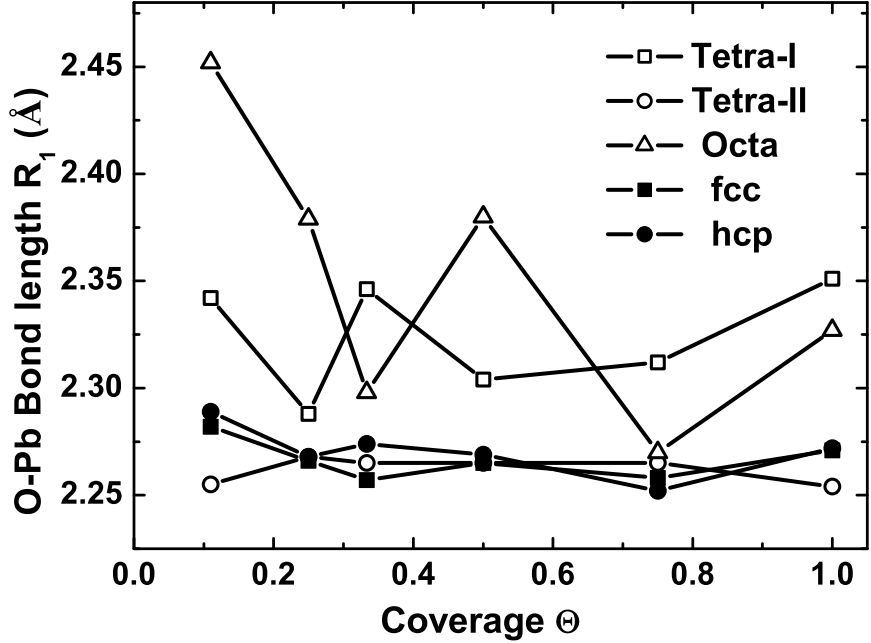}
\end{center}
\caption{The O-Pb bond length as a function of coverage for
different on-surfce and subsurface adsorption sites.}
\end{figure}%
The O-Pb bond length $R_{1}$ for the fcc and hcp on-surface adsorption sites
is shown in Fig. 5 as a function of the oxygen coverage $\Theta$. One can see
that for both fcc and hcp adsorption, the O-Pb bond length varies around 2.26
\AA very little with increasing $\Theta$. In particular, the calculated
results of $R_{1}$ by using the same $p(2\times2)$ surface model varies only
within an amplitude of 0.01 \AA (0.02 \AA ) for fcc (hcp) site. The short bond
length $R_{1}$ implies a strong interaction between O and Pb atoms. Note that
the value of $R_{1}$ in the fcc adsorption site is a little shorter than in
the hcp site, which is consistent with the fact that the fcc site is more
stable than the hcp site for on-surface adsorption.%

\begin{figure}[tbp]
\begin{center}
\includegraphics[width=0.4\linewidth]{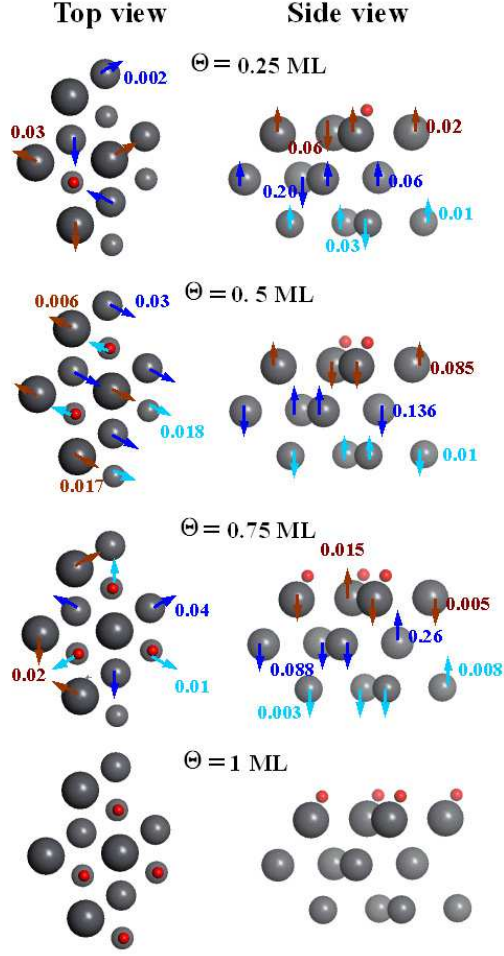}
\end{center}
\caption
{(Color online) Different structures of oxygen adsorbed in the fcc site on Pb(111) surface. The left part shows the top view of the three outmost Pb layers.
The right-hand part shows a side view of the vertical positions of the ions, cut in the plane along the $[1\bar
{2}1]$ direction. Note, the gray balls with different scales,
represent Pb atoms on the different layers (the lager is the outer), and the small red balls represent the oxygen atoms.
The arrows (not to scale) indicate the directions of the atomic displacements,
with different colors standing for different layers. The numbers that refer to the arrows are given in angstrom.}
\end{figure}%
Upon on-surface oxygen adsorption, the Pb atoms on the three outmost layers
exhibit lateral and vertical displacements, which are plotted in Fig. 6 for
energetically stable fcc adsorption at different coverages. From the left
panel (atomic top view) of Fig. 6, one can see that Pb atoms bonded to oxygen
in top layer always move outwards from chemisorbed oxygen for different
coverages with the same p$(2\times2)$ surface model. The values of atomic
movements are also indicated in the figure. The right panel (side view)\ of
Fig. 6 depicts the vertical movements of Pb atoms from the center of mass of
each layer. One prominent feature is that the oxygen at various coverages
(except for the case $\Theta$=1.0) we considered causes a most noticeable
rumpling of the second Pb layer, which can be seen by large value of vertical
movement of the second Pb layer.%

\begin{figure}[tbp]
\begin{center}
\includegraphics[width=0.7\linewidth]{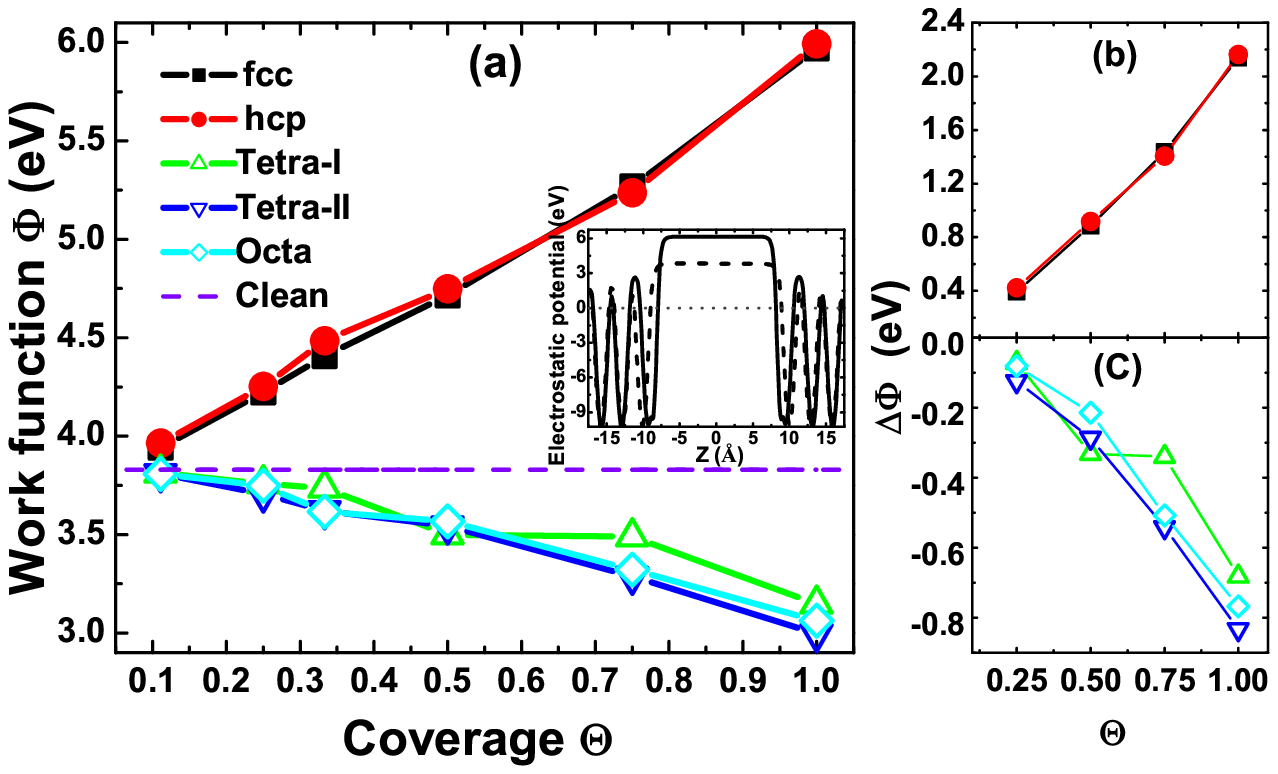}
\end{center}
\caption{(Color online) (a) The calculated work function $\Phi$ as a function
of $\Theta$ for different on-surface and subsurface adsorption sites. The
dashed line shows the value of $\Phi$ for clean Pb(111) substrate. (b) The
change $\Delta\Phi$ in the work function by fcc and hcp on-surface O
adsorption versus the coverage. (c) The quantity $\Delta\Phi$ for tetra-I,
tetra-II, and octa subsurface adsorption sites. As an example, the inset in figure (a) shows the planar-averaged electrostatic
potential of clean (dashed curve) and O-adsorbed (solid curve with $\Theta
$=$1.0$) Pb(111) slab, with the Fermi energies for both cases set to be zero.}
\end{figure}%
We turn now to analyze the electronic properties of the O/Pb(111) system by
first considering the work function $\Phi$ at different coverages and its
change $\Delta\Phi$ with respect to the clean Pb(111) surface, both of which
were illustrated in Fig. 7 and summaized in Table II. From Fig. 7(a), it can
be seen that the work function slightly but steadily increases with O coverage
for both fcc and hcp on-surface adsorption sites. Comparing with the other TM
surfaces such as Ag(111), we notice that the varying amplitude of $\Phi$ in
the whole coverages at the present Pb(111) surface is much smaller. In fact,
from Fig. 7(b) one can see that the value of $\Delta\Phi$ varies from about
0.4 eV (at $\Theta$=0.25) to about 2.1 eV (at $\Theta$=1.0) for both fcc and
hcp on-surface adsorption. Whereas, for O/Ag(111) system, $\Delta\Phi$ changes
from about 1.2 eV at $\Theta$=0.25 to 4.0 eV at at $\Theta$=1.0
\cite{LiWX-2002}. On the other side, our results of $\Delta\Phi$ versus
$\Theta$ can be comparable with those for the other O-adsorbed $sp$ simple
metal surfaces such as Al(111) \cite{Al(111)-2} and Mg(0001) \cite{Mg(0001)}.
Also from Fig. 7 it can be seen that the amplitude of $\Delta\Phi$ for the hcp
adsorption site is a little more prominent than that of the fcc site.%

\begin{figure}[tbp]
\begin{center}
\includegraphics[width=1.0\linewidth]{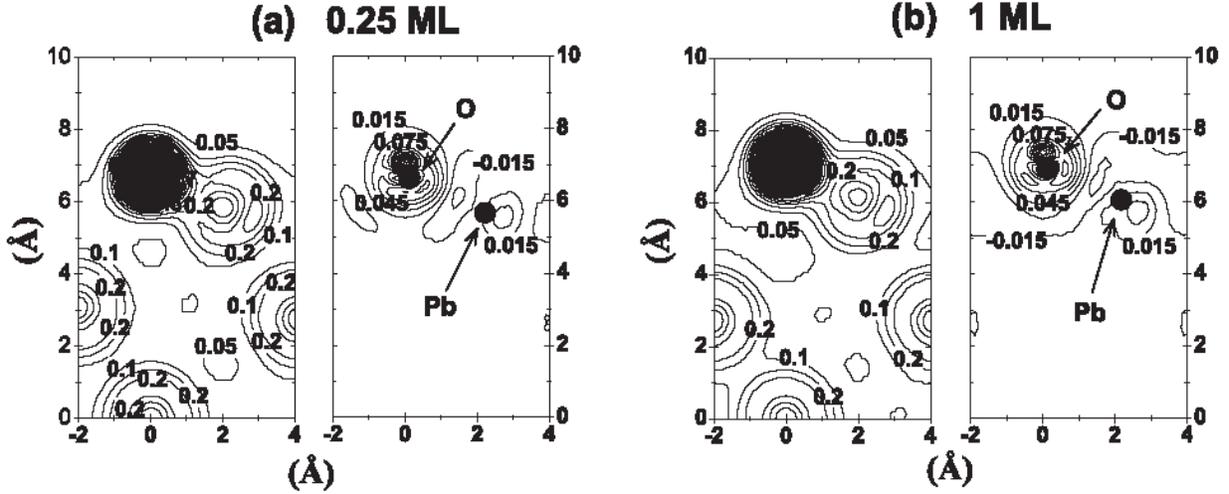}
\end{center}
\caption{Contour plots of the total valence charge density $n(\mathbf{r})$
(left panel) and the charge density difference $\Delta n(\mathbf{r})$ (right
panel) for the on-surface O/Pb(111) slabs with the coverage (a) $\Theta
=0.25$ and (b)
$\Theta=1.0$. The contour plane is parallel with the [1\={2}%
1] direction and is perpendicular to the Pb(111) surface.}
\end{figure}%
To gain more insight into the nature of oxygen chemisorption on Pb(111)
surface, we now analyze our results by means of the charge density difference
$\Delta n(\mathbf{r})$ before and after O adsorption, and the orbital-resolved
PDOS. Here the density difference $\Delta n(\mathbf{r})$\ was obtained by
subtracting the electron densities of noninteracting component systems,
$n^{\text{Pb(111)}}(\mathbf{r})+n^{\text{O}}(\mathbf{r})$, from the density
$n(\mathbf{r})$ of the O/Pb(111) system, while retaining the atomic positions
of the component systems at the same location as in O/Pb(111). Figures
8(a)-(b) (right panel in each figure) present the contour plots of $\Delta
n(\mathbf{r})$ for $\Theta$=0.25 and $\Theta$=1.0, respectively. The total
charge densities are also shown in the left panels. Such contour plots can
provide physical insight into the nature of the chemical bonding. It reveals
that the charge redistribution mainly occurs at the surface and involves the O
adatom and the topmost Pb atoms. It is apparent that upon adsorption electrons
flow into the O-2$p$ orbitals. The influence of the adsorbed surface is
rapidly screened out on going into the bulk. In fact, one can see that the
bonding character of the inner Pb layers (from the second layer for $\Theta
$=0.25 and from the third layer for $\Theta$=1.0) is essentially identical to
the bulk case, which is typically metallic with a fairly constant charge
density between the atoms with slight directional bonding along the body
diagonals. This is also emphasized by the density difference contour plots in
Fig. 8 which reveals only very small changes in the interior of the nine-layer
Pb slab. At both low ($\Theta$=0.25) and high ($\Theta$=1.0) coverages, one
can see from Fig. 8 that the second-layer Pb atom and especially the surface
Pb atom show remarkable changes in the charge density, reflecting the strong
influence of the neighboring on-surface O atoms. It is also apparent in Fig. 8
that the bonding between O and surface Pb atoms is largely ionic in nature,
which is is evident from the charge accumulation spherically centered on the O
atom with radius $\sim$1.35 \AA (comparable with 1.4 \AA of ionic radius of
O$^{-2}$) and the charge depletion between O and surface Pb atoms. The charge
accumulation on the surface and charge depletion below the surface will result
in the formation of a dipole moment (in line with the usual surface dipole
layer) which tends to increase the work function as compared to that for the
clean Pb(111) surface, as discussed above. Obviously, the charge transforms
solely onto the O-$2p$ orbitals. The shape of the density difference contour
is similar for low and high coverages, indicating that the O-Pb bonding nature
is keeps invariant when increasing the oxygen coverage.%

\begin{figure}[tbp]
\begin{center}
\includegraphics[width=0.7\linewidth]{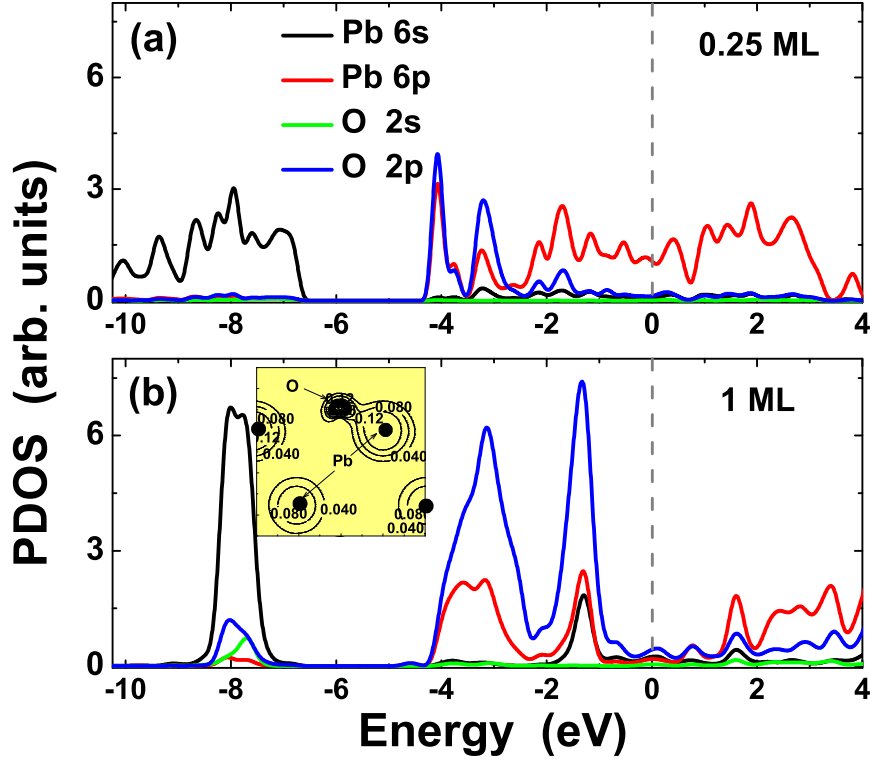}
\end{center}
\caption{(Color online) Orbital-resolved partial DOS for the surface Pb layer
and the on-surface O adatom (fcc site) with the coverage (a)  $\Theta=0.25$
and (b) $\Theta
=1.0$. The inset in (b) depicts the band charge density around the energy E=-8 eV.}
\end{figure}%
Figures 9(a) and (b) show the orbital-resolved PDOS for the on-surface
O$_{\text{fcc}}$ layer and the topmost Pb layer at $\Theta$=0.25 and $\Theta
$=1.0, respectively. At low coverage ($\Theta$=0.25), the left-side peak
[around $-$4.0 eV in Fig. 9(a)] of O-2$p$ states represents the degenerate
2$p_{x}$ and 2$p_{y}$ states, which hybridize with the degenerate 6$p_{x}%
$/6$p_{y}$ states of the outmost Pb atoms. The right-side peak [around $-$3.2
eV in Fig. 9(a)] of O-2$p$ states represents the mixed $p_{x,y}$ and $p_{z}$
states (with $p_{z}$ orbital a little lower in energy than $p_{x,y}$
orbitals), which hybridize with Pb-6$p_{x,y}$ and -6$p_{z}$ states. The
coupling between the Pb-6$s$ orbital and the O-2$p$ orbitals is negligibly
small at low coverage. In addition, it shows in Fig. 9(a) that the metallic
and bonding nature of surface Pb(111) layer does not change by low coverage of
oxygen. This can be seen by the facts: (i) The PDOS at $E_{F}$ for the topmost
Pb layer is large, comparable with that for clean Pb(111) surface; (ii) The
surface Pb-2$s$ and -2$p$ states change very little upon oxygen adsorption. In
addition, the previous analysis of the atomic geometry also showed that the
change in the first-second interlayer relaxation ($\Delta_{12}$) upon oxygen
adsorption is relatively small at low coverage, indicating the metallic
bonding between the first and second Pb layers preserves well. With increasing
oxygen coverage, the prominent changes involving the O-Pb interaction and the
surface Pb bonding occur, which can be seen from Fig. 9(b) for $\Theta$=1.0.
First, the density of the surface-layer Pb-6$p$ states at the Fermi level is
significantly suppressed. A more clear illustration can be found in Fig. 10.
Second, the energy distribution of the surface-layer Pb-6$s$ state becomes
very narrow compared to its original broad-dispersion feature in the clean
surface, which indicates that unlike the low-coverage case, the surface Pb $s$
state at high coverage becomes very isolated and dense. The band charge
density around the energy $E$=$-8$ eV is plotted in the inset in Fig. 9(b),
which shows that at this energy interval, the surface Pb atoms have more and
dense charge compared to the second-layer Pb atoms. Third, the energy distance
between the two O-2$p$ (or their hybridized Pb-6$p$) peaks increases from 0.8
eV at $\Theta$=0.25 to 2.0 eV at $\Theta$=1.0. Also, these two peaks become
more broadened by the increase of $\Theta$. In addition, one can see from Fig.
9(b) that at high oxygen coverage, there appears an increasing hybridization
between the O-2$p$ and the top-layer Pb-6$s$ states around $E$=$-$8.0 eV. In a
similar way, a hybridization between these two kinds of states at high
energies around $E$=$-$1.3 eV also becomes obvious.
\begin{figure}[tbp]
\begin{center}
\includegraphics[width=0.7\linewidth]{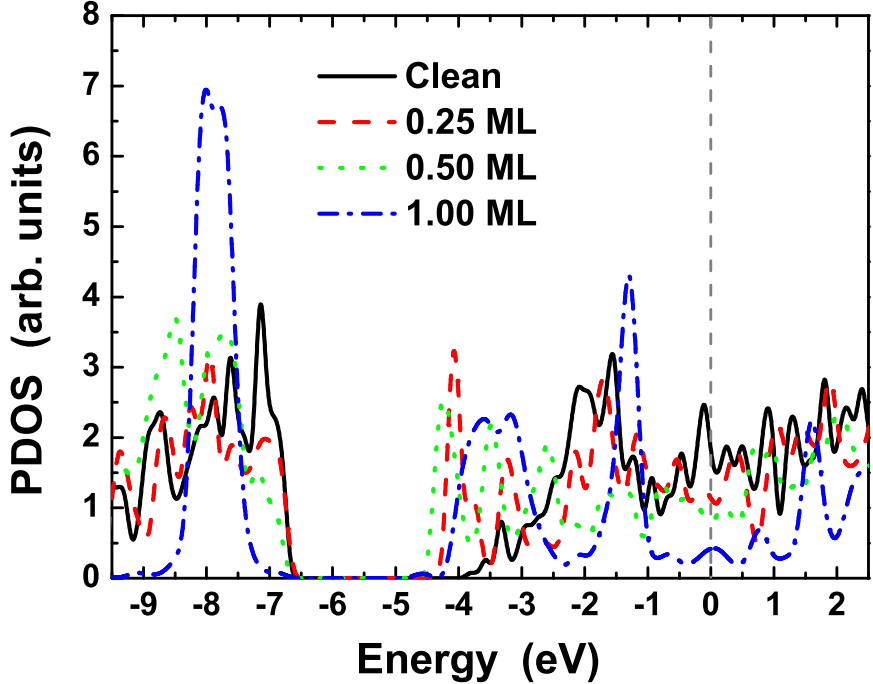}
\end{center}
\caption
{(Color online) Total DOS of the on-surafec O adatoms (fcc site) and the
surface-layer Pb atoms for different values of the coverage $\Theta$.}
\end{figure}%
For further illustration, we plot in Fig. 10 the total DOS (a sum over
O$_{\text{fcc}}$ and topmost Pb layers) for different coverages calculated by
the p$(2\times2)$ surface model. One can see that the density at the Fermi
energy monotonically decreases with increasing the coverage, and the metallic
$p$ bands of the topmost Pb gradually evolve into the insulating bands due to
its ionic coupling with the O-2$p$ orbitals. Also the original broad
distribution of the surface Pb-2$s$ state is increasingly narrowed with
coverage. Therefore, Fig. 10 clearly the tendency to the insulating surface by
the increase of the coverage.

\section{THE PURE SUBSURFACE ADSORPTION}

As we did in Sec. IV, we continue to study the (incorporated) subsurface
adsorption of atomic oxygen without the presence of on-surface oxygen, namely,
pure subsurface. For oxygen occupation in the subsurface region there are
three different high-symmetry sites. The octahedral site (henceforth octa)
lies just underneath the fcc on-surface site, and one tetrahedral site
(tetra-I) lies below the hcp on-surface site. A second tetrahedral site
(tetra-II) is located directly below a first layer metal atom. The coordinates
of these subsurface sites have been schematically shown in Fig. 1. We
performed calculations for oxygen in these different sites for a wide range of
coverages, i.e., from 0.11 ML to 1.0 ML. Note that we focus on adsorption
immediately below the first Pb layer as we find that oxygen adsorption deep in
the bulk is less favorable in every case. The cases of simultaneous subsurface
and on-surface oxygen adsorption will be studied in a forthcoming paper. In
the following, the subsurface energetics, the atomic structure, and the
electronic properties are discussed in detail.

The calculated binding energies $E_{b}$ for O$_{\text{tetra-I}}$,
O$_{\text{tetra-II}}$, and O$_{\text{octa}}$, with respect to atomic oxygen,
are plotted in Fig. 11
\begin{figure}[tbp]
\begin{center}
\includegraphics[width=0.7\linewidth]{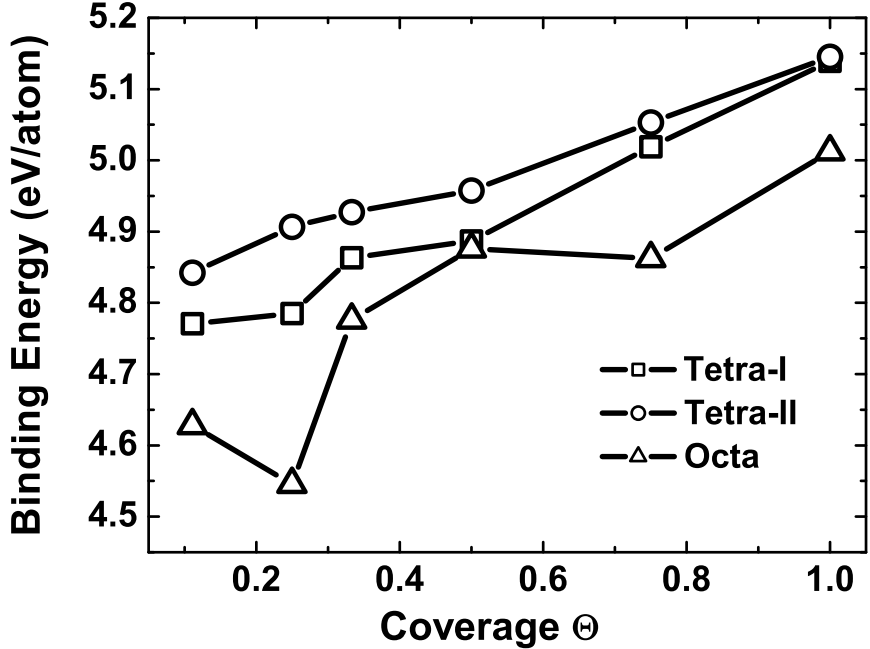}
\end{center}
\caption{Binding energies $E_{b}%
$ for oxygen in the subsurface tetra-I, tetra-II and octa adsorption sites, as functions of the O coverage.}
\end{figure}
and summarized in Table II. It can be seen from Table II that the most
preferred adsorption site is tetra-II site for the whole coverage range and
among all the on-surface and subsurface adsorption sites considered. This is
different from the other systems such as O/Ag(111), O/Cu(111), and O/Mg(0001).
For O$_{\text{tetra-II}}$ (open circles in Fig. 11) or O$_{\text{tetra-I}}$
(open squares), the binding energy increases slightly but steadily with O
coverage, which is similar to the case of on-surface adsorption. For
O$_{\text{octa}}$ adsorption (open triangles), it reveals in Fig. 11 that the
binding energy decreases when the coverage $\Theta$ varies from 0.11 to 0.25,
indicating a repulsive interaction between the O$_{\text{octa}}$ atoms;
however, it increases again from $\Theta$=0.33 to 1.0, so that the effective
interaction between the O$_{\text{octa}}$ atoms in this coverage range is
attractive. Despite these different dependence on coverage, we note that the
overall variation in the magnitude of the binding energy as a function of the
coverage is rather small for the three subsurface sites, namely, 0.33 eV, 0.51
eV, and 0.50 eV for O$_{\text{tetra-II}}$,\ O$_{\text{tetra-I}}$, and
O$_{\text{octa}}$ atoms, respectively. Compared to the on-surface adsorption,
the subsurface adsorption is notably more favorable in the whole range of
coverages considered. In fact, the binding energy difference $\Delta E_{b}$
between the most energetically stable on-surface O$_{\text{fcc}}$ and
subsurface O$_{\text{tetra-II}}$ atoms increases from about\ 0.1 eV to 0.2 eV
with increasing the coverage $\Theta$ from 0.11 to 1.0. The result that the
subsurface adsorption is more stable than the on-surface adsorption has also
been found in O/Mg(0001) system \cite{Mg(0001)}, but with the different
tendency with respect to the variation of the O coverage. In the O/Mg(0001)
system it was found that the value of $\Delta E_{b}$ decreases from about\ 0.4
eV to less than 0.1 eV with the increasing coverage $\Theta$ from 0.0625 to
1.0 \cite{Mg(0001)}. \ \

\begin{table}[th]
\caption{The calculated interlayer relaxation $\Delta_{ij}$ (\%) and O-Pb bond
length $R_{1}$ (in \AA ) for different oxygen\ coverages of subsurface
adsorption. }
\begin{tabular}
[c]{cccccccccc}\hline\hline
Coverage & \multicolumn{3}{c}{$R_{1}$ (\AA ) \ } & \multicolumn{3}{c}{$\Delta
_{12}$ (\%) \ } & \multicolumn{3}{c}{$\Delta_{23}$ (\%)}\\
$\Theta$ & tetra-I & tetra-II \  & octa \  & tetra-I & tetra-II \  & octa \  &
tetra-I & tetra-II \  & octa \ \\\hline
0.11 & 2.342 & 2.255 & 2.452 \  & -3.238 & -2.058 & -3.065 \  & 2.875 &
1.516 & 1.703 \ \\
0.25 & 2.288 & 2.268 & 2.379 \  & 0.129 & 0.757 & 0.037 \  & 3.773 & 2.577 &
0.436 \ \\
0.33\  & 2.346 & 2.265 & 2.497 \  & -1.405 & 2.205 & 0.373 \  & 6.139 &
2.607 & 1.840 \ \\
0.50 & 2.304 & 2.265 & 2.380 \  & 12.22 & 6.547 & 13.77 \  & 3.190 & 3.301 &
2.557 \ \\
0.75 & 2.312 & 2.265 & 2.270 \  & 15.16 & 13.92 & 13.55 \  & 6.022 & 4.809 &
-1.861 \ \\
1.00 & 2.351 & 2.254 & 2.327 \  & 25.18 & 21.80 & 22.25 \  & 1.407 & 5.420 &
-3.988 \ \\\hline\hline
\end{tabular}
\end{table}

Table IV presents the results of the O-Pb bond length $R_{1}$ and the
interlayer relaxations ($\Delta_{12}$ and $\Delta_{23}$) in a wide range of
coverages for O$_{\text{tetra-I}}$, O$_{\text{tetra-II}}$, and O$_{\text{octa}%
}$ atoms.
\begin{figure}[tbp]
\begin{center}
\includegraphics[width=0.5\linewidth]{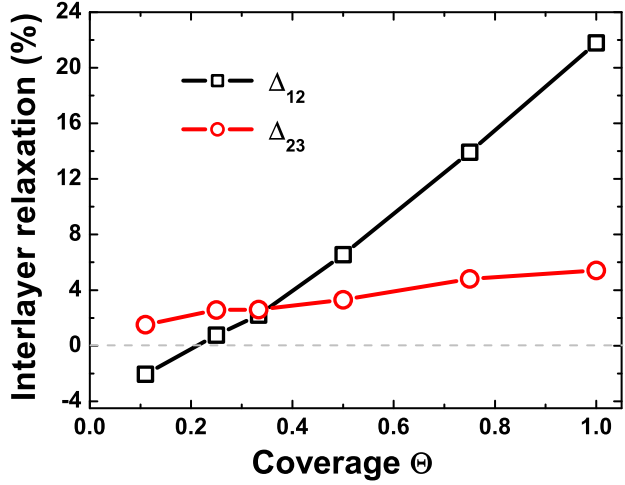}
\end{center}
\caption
{Interlayer relaxations as functions of coverage for oxygen in subsurface tetra-II site.}
\end{figure}%
For more clear illustration, the first ($\Delta_{12}$) and second
($\Delta_{23}$) interlayer relaxations with O in the most stable tetra-II
subsurface site are also plotted in Fig. 12 as functions of $\Theta$. The O-Pb
bond length for subsurface adsorption with different coverages has been
plotted in Fig. 5 together with the case of on-surface adsorption. One can see
from Fig. 12 that the first interlayer relaxation $\Delta_{12}$ increases from
$-$5\% to about 22\% (in a linear form) as a function of oxygen coverage,
while the second interlayer relaxation $\Delta_{23}$ is much insensitive to
the variation of the coverage compared to $\Delta_{12}$. From the variation of
the O-Pb bond length $R_{1}$ as a function of $\Theta$ (Fig. 5), one can see
that the value of $R_{1}$ for O$_{\text{tetra-II}}$\ atoms keeps the most
shortest and stable (against $\Theta$) compared to the other various kinds of
on-surface and subsurface adsorption sites. Whereas, for the octa\ and tetra-I
adsorption sites, it shows in Fig. 5 that the O-Pb bond length oscillates in a
large amplitude with increasing $\Theta$.

We turn now to study the electronic property of the subsurface O/Pb(111)
system by first considering the work function $\Phi$ at different coverages
and its change $\Delta\Phi$ with respect to the clean Pb(111) surface, both of
which were illustrated in Fig. 7 and summarized in Table II. From the results
shown in Fig. 7(a), it can be seen that the work function slightly but
steadily decreases with the coverage for three subsurface adsorption sites.
The overall variation in the magnitude of the $\Delta\Phi$ is rather small,
namely, $-$0.8 eV, $-$0.66 eV, and $-$0.73 eV for O$_{\text{tetra-II}}$,
O$_{\text{tetra-I}}$, and O$_{\text{octa}}$, respectively. These results of
$\Delta\Phi$ versus $\Theta$ can be comparable with those in O/Al(111) system
\cite{Al(111)-2}, in which relatively small work-function change was observed
despite the strong electron transfer between adsorbate and Al atoms, and was
associated with the small adsorption distance for the O species
\cite{Lang1983}. Also from Figs. 7(a)-(b) it can be seen that the amplitude of
$\Delta\Phi$ for the tetra-II adsorption is most prominent among the three
subsurface sties.%

\begin{figure}[tbp]
\begin{center}
\includegraphics[width=1.0\linewidth]{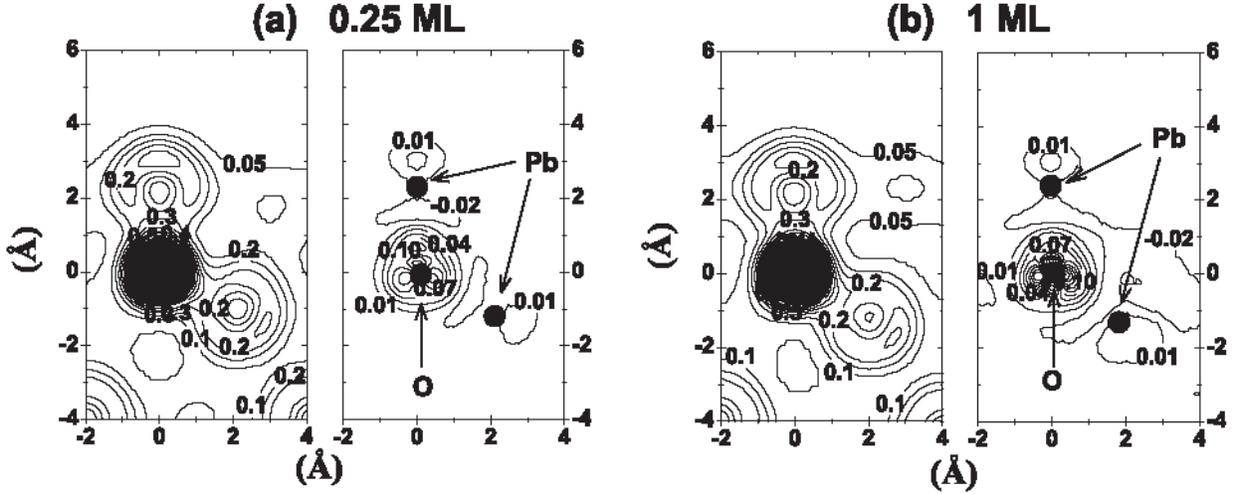}
\end{center}
\caption
{Contour plots of the total charge density (left panel) and the charge density difference (right
panel) for the subsurface O$_{\text{tetra-II}}%
$/Pb(111) slabs with the coverage (a) $\Theta=0.25$ and (b)
$\Theta=1.0$. The contour plane is parallel with the [1\={2}%
1] direction and is perpendicular to the Pb(111) surface.}
\end{figure}%
To gain more insight into the nature of the O subsurface adsorption, we now
analyze our results by means of the charge density difference $\Delta
n(\mathbf{r})$. Figures 13(a)-(b) (right panel in each figure) present the
contour plots of $\Delta n(\mathbf{r})$ for $\Theta$=0.25 and $\Theta$=1.0,
respectively, for the tetra-II subsurface adsorption. The total charge
densities are also shown in the left panels. It reveals that the charge
redistribution mainly involves the subsurface O and the neighboring outmost
two Pb layers. The electrons flows upon adsorption from the first- and
second-layer Pb atoms to the O-2$p$ orbitals. It is apparent that the adsorbed
O-2$p$ orbitals are not the same as those of the free O atom but they are
polarized along the three Pb-O axis. This polarization is reasonable because
the Pb atoms are positively charged. Also one can see from Fig. 13 that the
bonding character of the inner Pb layers (from the third layer for the whole
coverages considered) is essentially identical to the bulk case, while the
first- and second-layer Pb atoms show remarkable changes in the charge
density, reflecting the strong influence of the neighboring subsurface O atom.
As with the on-surface adsorption, since there is no electron density
accumulation in the O-Pb bondng regions, we obtain an ionic-like bonding
picture for the subsurface O atom, with charge transfer and polarization. The
subsurface accumulation and the on-surface depletion in the charge density
would result in the formation of a dipole moment (anti-paralleling with the
usual surface dipole layer) which tends to reduce the work function as
compared to that for the clean Pb(111) surface, as shown in Fig. 7.%

\begin{figure}[tbp]
\begin{center}
\includegraphics[width=0.7\linewidth]{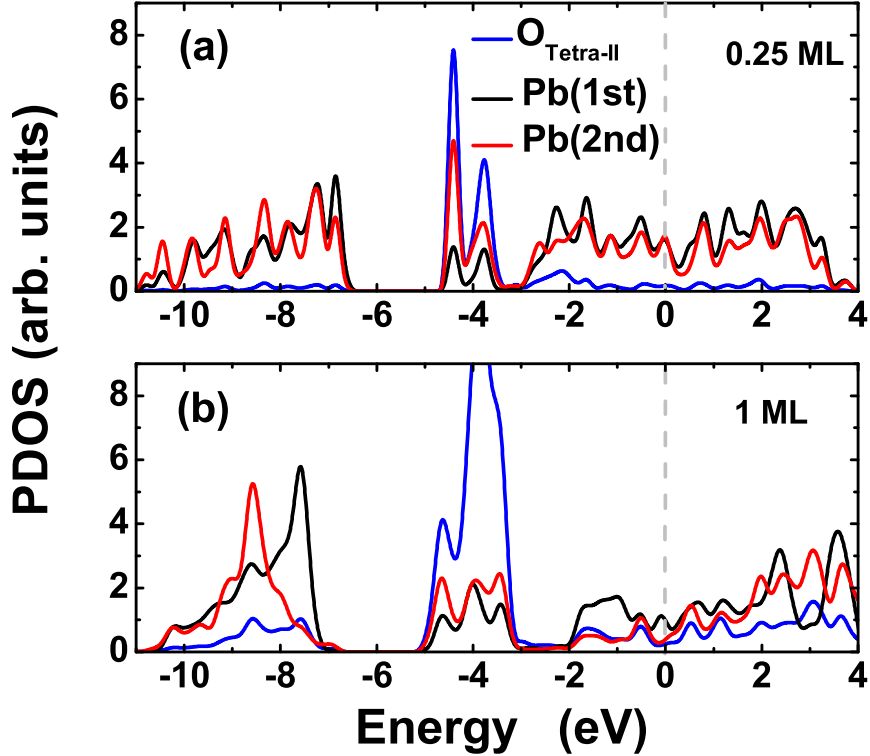}
\end{center}
\caption{(Color online) Layer-projected DOS for the O$_{\text{tetra-II}}$
subsurface layer and the outmost two Pb(111) layers.}
\end{figure}%
Figures 14(a)-(b) shows the PDOS for the subsurface O$_{\text{tetra-II}}$
layer and the first and second Pb(111) layers at coverages $\Theta$=0.25 and
$\Theta$=1.0, respectively. Similarly to on-surface oxygen adsorption, the
interaction between subsurface oxygen and lead is mainly via hybridization of
the O-2$p$ and Pb-6$sp$ states. During this hybridization, however, the PDOS
energy distribution of two Pb layers and of the 6$s$ and 6$p$ orbitals of the
same Pb layer are prominently different. In fact, it can be seen from Fig.
14(a) that at low coverage, the O-2$p$ states mainly hybridize with the
Pb-6$p$ states, while the interaction between the O-2$p$ states and the
Pb-6$s$ states is negligibly small. The O-2$p$ states couple to 6$p$ states of
the second Pb layer more strongly than those of the first Pb layer, since the
two peaks around $-4$ eV (below the Fermi energy) in Fig. 14(a) mainly consist
of 2$p$ states of O$_{\text{tetra-II}}$ atoms and 6$p$ states of the second Pb
layer, while the contribution from the first-layer Pb atoms is small. The
reason is simply due to the fact that at low coverage, every
O$_{\text{tetra-II}}$ atom is bonded to its four neighboring Pb atoms, three
of which come from the second Pb layer, while only one comes from the first Pb
layer. At as high coverage as $\Theta$= 1.0 [Fig. 14(b)], there are the
following new features: (i) The PDOS shifts downwards in energy and a large
spectra weight near the Fermi energy (from below) becomes very small, both of
which obviously will lead to the increase in the adsorption energy; (ii) The
coupling between the O-2$p$ and Pb-2$s$ states comes to play an important role
in the O-Pb bond. Due to the difference in the coordinates of the neighboring
Pb atoms, this coupling is asymmetric for the first- and second-layer Pb
atoms, thus giving rise to a weight splitting in the 6$s$ PDOS of the two Pb
layers as shown in Fig. 14(b); (iii) With increase of O$_{\text{tetra-II}}$
coverage, the coupling between the O-2$p$ states and first-layer Pb-2$p$
states is enhanced, which is revealed by the peaks around $-$4.0 eV (below the
Fermi energy) in Fig. 14(b); (iv) Compared to the high-coverage on-surface
adsorption [Fig. 9(b)], the narrowing effect of the Pb-2$s$ DOS by the
coupling with O-2$p$ states is not prominent so much.

\section{The energy barrier for atomic oxygen diffusion}

The diffusion of the atomic O after the on-surface dissociation of the O$_{2}$
molecule is an elementary process during the whole surface oxidation process,
which is still in debate for many systems regarding the existence of
\textquotedblleft hot\textquotedblright\ atoms with transient mobility upon
O$_{2}$ dissociation \cite{Win1996,Har1981,Bru1993}. Also, the oxygen
diffusion plays a key role in understanding many important catalytic
reactions, such as oxidation of CO and hydrocarbons in the catalytic treatment
of the automotive exhaust gases. In this section, by using the DFT total
energy calculation, we report our numerical results of\ the energy barriers
for atomic O diffusion and penetration in the O/Pb(111) system.%

\begin{figure}[tbp]
\begin{center}
\includegraphics[width=1.0\linewidth]{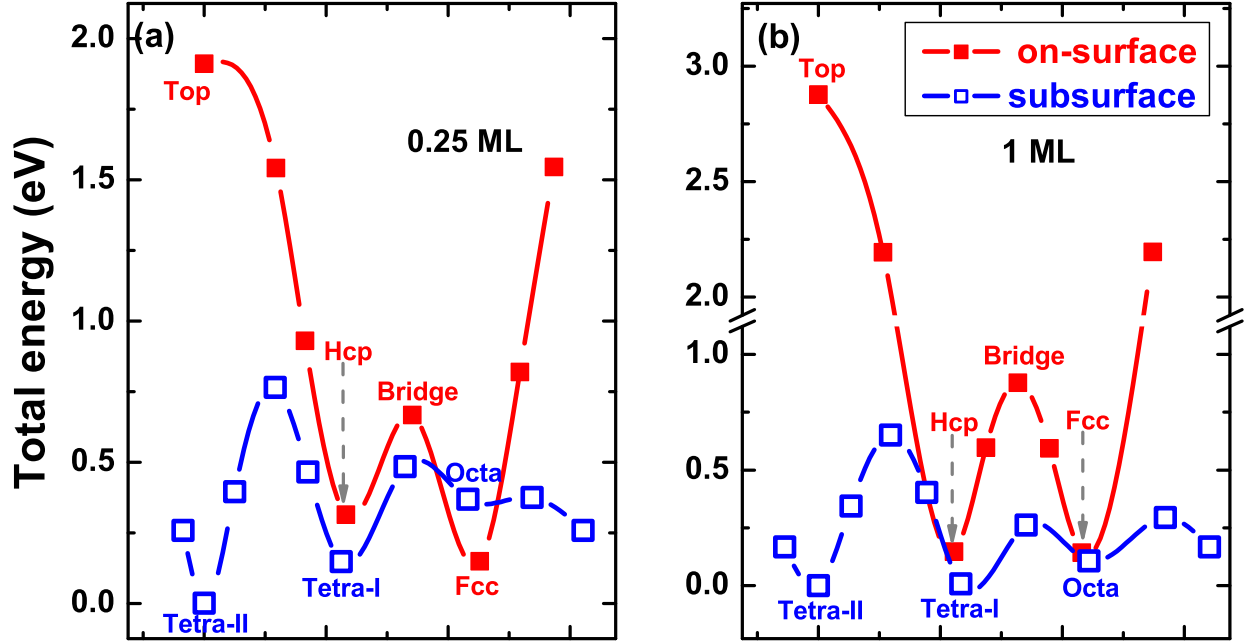}
\end{center}
\caption{(Color online) Total energy for an oxygen atom carrying out the
on-surface (filled squares) or the subsurface (open squares) diffusion on Pb(111)
face at the O coverage of (a) $\Theta=0.25$ and (b) $\Theta=1.0$. Here the
total energy is given in reference to the most stable subsurface tetra-II
occupation.}
\end{figure}%
Using the nudged elastic band (NEB) method \cite{mills,mills-2,Alatalo}, which
is capable of finding saddle points and minimum energy paths on complicated
potential surfaces, we have calculated the on-surface and subsurface
diffusion-path energetics of atomic oxygen. The results are shown in Fig.
15(a) for low ($\Theta$=0.25) coverage and in Fig. 15(b) for high ($\Theta
$=1.0) coverage. Note that in each figure [Fig. 15(a) or (b)] the number of
atoms keeps invariant in the on-surface and subsurface calculations. Thus it
also reveals in Fig. 15 the relative stability of O adsorption among various
on-surface and subsurface sites and the corresponding oxygen binding energy
differences. For the on-surface adsorption, our calculated diffusion barrier
from fcc to hcp site is 0.52 eV at $\Theta$=0.25 and 0.74 eV at $\Theta$=1.0.
The hcp site is less stable than the fcc site within the coverages $0<$
$\Theta\leq1$, although the binding energy difference between O$_{\text{fcc}}$
and O$_{\text{hcp}}$ decreases rapidly when increasing $\Theta$ (see Fig. 3).
Thus the on-surface diffusion barrier from hcp to fcc site presents an
activation barrier with the value of 0.35 eV at $\Theta$=0.25 and of 0.73 eV
at $\Theta$=1.0. For the subsurface adsorption, Fig. 15 shows that all of the
three sites we considered, i.e., the tetra-I, the tetra-II and the octa sites
present the local energy minimum along the oxygen subsurface diffusion path.
The subsurface diffusion barrier from tetra-II to tetra-I is 0.76 eV at
$\Theta$=0.25 and 0.64 eV at $\Theta$=1.0. For the diffusion from the tetra-I
site to the octa site, the calculated energy barrier is 0.28 eV at $\Theta
$=0.25 and 0.25 eV at $\Theta$=1.0. For the diffusion from the tetra-II to
octa site, the calculated barrier is 0.37 eV at $\Theta$=0.25 and 0.29 eV at
$\Theta$=1.0. The octa site is less stable than the tetra-I site, while the
tetra-I site is less stable than the tetra-II site. Thus the subsurface
diffusion barrier from tetra-I to tetra-II site presents an activation barrier
with the value of 0.62 eV at $\Theta$=0.25 and of 0.64 eV at $\Theta$=1.0. In
the same way, the activation barrier for octa$\rightarrow$tetra-I diffusion is
0.11 eV at $\Theta$=0.25 and 0.16 eV at $\Theta$=1.0, while the activation
barrier for octa$\rightarrow$tetra-II diffusion is 0.01 eV at $\Theta$=0.25
and 0.18 eV at $\Theta$=1.0.%

\begin{figure}[tbp]
\begin{center}
\includegraphics[width=0.7\linewidth]{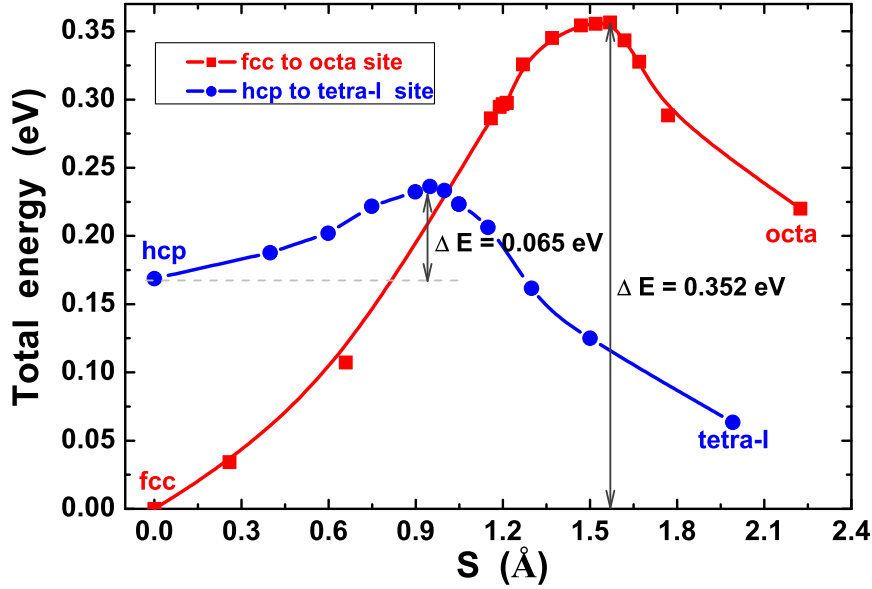}
\end{center}
\caption{(Color online) Total energy for an oxygen atom carrying out the
penetration from the on-surface fcc to the subsurface octa site (filled
squares) or from the on-surface hcp to the subsurface tetra-I site (filled
circles) of Pb(111) lattice. Here the O coverage is set $\Theta=0.25$ and the
total energy is given in reference to the on-surface fcc occupation. $s$
coordinate indicates the penetration distance of the O atom with respect to
the initial on-surface fcc or hcp adsorption site.}
\end{figure}%
To investigate subsurface oxygen in more detail with regard to the energetics
of its formation, we now study the penetration process of oxygen atom. Note
that although the subsurface tetra-II site is most stable, the direct oxygen
penetration into this site from the on-surface adsorption without by-passing
the other subsurface sites is very unfavorable, since this site is located
below a surface Pb atom. Therefore, here we only consider the penetration of O
atom through the outmost Pb layer from the on-surface fcc (hcp) to the
neighboring subsurface octa (tetra-I) site at the coverage $\Theta$=0.25. For
this purpose, we fully relaxed the first three Pb layers, and restrain the
oxygen atom when approaching to the first Pb layer step by step to search for
the transition state with high energy.\ The calculated penetration paths and
the energy barriers from the on-surface fcc to subsurface octa site and from
the on-surface hcp to subsurface tetra-I site are shown in Fig. 16, in which
the $s$ coordinate indicates the penetration distance of the O atom with
respect to the initial on-surface fcc or hcp adsorption site. The calculated
penetration barrier from on-surface fcc to subsurface octa site is 0.352 eV.
The transition state, i.e., the atomic geometry of the energy maximum in the
penetration path, corresponds to the oxygen atom 0.1 \r{A}above the surface Pb
layer. For penetration from the on-surface hcp to the subsurface tetra-I site,
the calculated activation barrier is as small as 0.065 eV, which implies the
most favorable path for the oxygen penetration. The transition state
corresponds to the oxygen atom 0.4 \r{A} below the surface Pb layer. Comparing
to the energy barriers for the O on-surface diffusion, one can see that the
oxygen atoms can easily intrude into the subsurface and bond with the
second-layer Pb atoms rather than searching for a more stable on-surface site
(fcc site) to settle down. Even though an oxygen atom settles down at the
stable on-surface fcc site, it also can penetrate into the subsurface
stimulated by a slight perturbation.

\section{CONCLUSION}

In summary, we have systematically investigated the adsorption of atomic
oxygen on the Pb(111) surface and subsurface, as well as the energy barriers
for atomic O diffusion and penetration in these systems through
first-principles DFT-GGA calculations. We considered a wide range of coverage
using different surface models [i.e., p$(3\times3)$, p$(2\times2)$,
p$(\sqrt{3}\times\sqrt{3})$, and p$(1\times1)$\ surface unit cells] for
adsorption in the on-surface fcc and hcp sites as well as the subsurface
tetra-I, tetra-II, and octa sites. For the on-surface adsorption, the fcc site
is more stable than the hcp site for the whole coverage range considered. The
oxygen binding energy difference between these two sites decreases with
increasing coverage. In particular, at the coverage of 1 ML, the binding
energy of O$_{\text{hcp}}$ is almost identical from below to that of
O$_{\text{fcc}}$, implying a critical coverage for the stability conversion
between the on-surface fcc and hcp sites. The atomic geometry, the
work-function change, the charge density distribution, and the electronic
structure upon the O\ on-surface adsorption have also been studied, which
consistently show the fundamental influence by the ionic bonding between the O
atom and the first-layer Pb atoms with charge transfer from the latter to the
former. Remarkably, this influence in the energetics and atomic structure is
monotonically enhanced with increasing the O coverage, which is highly
interesting. For instance, the increase of the O binding energy for the fcc or
the hcp site with increasing the coverage implies the effective attraction
between the O adsorbates, which will make it favorable for the formation of
the oxygen island or cluster. Furthermore, we have found that with increasing
coverage, the electronic PDOS at the Fermi energy rapidly decreases, which
will further stabilize the system in accordance with the increase in binding energy.

For the purely subsurface O adsorption, the tetra-II site is more stable than
the tetra-I and octa sites for the whole coverage range considered. A similar
correlation between the binding energy and the coverage has been found for
O$_{\text{tetra-I}}$ and O$_{\text{tetra-II}}$, which also implies the
effective attraction between the adsorbed O atoms and the tendency to form the
oxygen island or cluster at subsurface sites. The other atomic and electronic
structural properties of the subsurface adsorption have also been
investigated. The observed increase of the first-second Pb interlayer
expansion, the decrease of the work-function change $\Delta\Phi$, and the
decrease of the PDOS at the Fermi energy as functions of the O coverage,
consistently show the stabilization of the ionic O-Pb bond with the charge
transfer from the first- and second-layer Pb atoms to the subsurface O atom.

Given the observation that the oxygen atom is more stable for the subsurface
than for the on-surface adsorption, we have calculated the on-surface and
subsurface diffusion-path energetics of O. The activation barrier for the
on-surface or the subsurface O diffusion becomes high when increasing the
coverage, which indicates the competition between the attractive interaction
from the adsorbate and the substrate and the repulsive force among the
adsorbates. In particular, we have shown that the activation barrier for the
penetration from the on-surface hcp to the subsurface tetra-I site is as low
as 0.065 eV, which indicates that the oxygen atoms can directly incorporate
into the lead (below the topmost Pb layer) right after on-surface O$_{2}$
dissociation at low coverage rather than nearly completion of a dense O adlayer.

\begin{acknowledgments}
This work was supported by the NSFC under grants Nos. 10604010 and 10544004.
\end{acknowledgments}

\end{document}